\documentclass[runningheads]{llncs}
\usepackage[T1]{fontenc}
\usepackage{graphicx} 
\usepackage{url}

\usepackage{amsmath}
\usepackage{amssymb}
\usepackage{bbm}
\usepackage{braket}
\usepackage{csquotes}
\usepackage{nicefrac}
\usepackage{tikz}
\usepackage{pgfplots}
\usepackage{wrapfig}
\usepackage{sidecap}
\usepackage{marvosym}

\usepackage{booktabs}

\pgfplotsset{compat=1.18} 

\newcommand{\bone}{\boldsymbol{1}}
\newcommand{\bx}{\boldsymbol{x}}

\newcommand{\bw}{\boldsymbol{w}}
\newcommand{\bp}{\boldsymbol{p}}
\newcommand{\bmu}{\boldsymbol{\mu}}
\newcommand{\bQ}{\boldsymbol{Q}}
\newcommand{\bM}{\boldsymbol{M}}
\newcommand{\bH}{\boldsymbol{H}}
\newcommand{\bE}{\boldsymbol{E}}
\newcommand{\bSig}{\boldsymbol{\Sigma}}

\newcommand{\mR}{\mathbb{R}}
\newcommand{\mP}{\mathbb{P}}
\newcommand{\mZ}{\mathbb{Z}}

\author{Sebastian Schlütter\inst{1}\and Tomislav Maras\inst{2}\and\\Alexander Dotterweich\inst{2}\and Nico Piatkowski\inst{3}(\Letter)%
}

\authorrunning{S. Schlütter et al.}

\institute{Mainz University of Applied Sciences, 55128 Mainz, Germany\and PricewaterhouseCoopers GmbH, 80636 München, Germany\and LAMARR Institute, Fraunhofer IAIS, 53757 Sankt Augustin, Germany}

\begin{document}

\title{Hot-Starting Quantum Portfolio Optimization}

\maketitle

\begingroup
  \renewcommand\thefootnote{}        
  \footnotetext{\Letter\;Corresponding author:
                \email{nico.piatkowski@iais.fraunhofer.de}}
  \addtocounter{footnote}{-1}        
\endgroup

\begin{abstract}
Combinatorial optimization with a smooth and convex objective function arises naturally in applications such as discrete mean-variance portfolio optimization, where assets must be traded in integer quantities. Although optimal solutions to the associated smooth problem can be computed efficiently, existing adiabatic quantum optimization methods cannot leverage this information. Moreover, while various warm-starting strategies have been proposed for gate-based quantum optimization, none of them explicitly integrate insights from the relaxed continuous solution into the QUBO formulation. In this work, a novel approach is introduced that restricts the search space to discrete solutions in the vicinity of the continuous optimum by constructing a compact Hilbert space, thereby reducing the number of required qubits. Experiments on software solvers and a D-Wave Advantage quantum annealer demonstrate that our method outperforms state-of-the-art techniques. 
\end{abstract}

\section{Introduction}

The integration of prior knowledge into mathematical models is essential. Examples include: (I) informed machine learning \cite{vR/etal/2023a}, where knowledge about physical laws allows for useful models even if training data is scarce. And (II) numerical optimization, where the convergence rate of gradient descent can be shown to depend on the squared Euclidean distance between the initial solution $\bx^{(0)}$ and the optimal solution $\bx^{*}$ \cite{Nesterov/2004a}. In the context of quantum computing, integrating prior knowledge about the solution is known as warm-start \cite{Beaulieu/Pham/2021a,Egger/etal/2021a}. An outstanding discussion of the topic can be found in \cite{Truger/etal/2024a}.

In case of \emph{Adiabatic Quantum Computation} (AQC) warm-start techniques act on the annealing schedule as the entry point since the initial state is built into the hardware and cannot be influenced---state preparation methods like \cite{Egger/etal/2021a} 
cannot be applied. Only a handful of practical methods is known, e.g. for training ML models to generate annealing schedules \cite{Chen/etal/2022a,Lin/etal/2022a} as well as techniques that rely on classical pre-processing \cite{deluis/etal/2022a,Truger/etal/2024a}. They can be seen as annealing equivalents to the techniques focusing on the parameter initialization for variational circuit, likewise pre-computing or generating annealing schedules instead of initial variational parameters.

\begin{SCfigure}[][t]
\centering
\begin{tikzpicture}
  \begin{axis}[
    xmin=13, xmax=18,
    ymin=21, ymax=26,
    xlabel={$\bx_1$},
    ylabel={$\bx_2$},
    axis equal image,
    xtick={13,14,15,16,17,18},
    ytick={21,22,23,24,25,26},
    grid=major,
    legend style={at={(0.97,0.03)},anchor=south east}, 
    legend cell align=left
  ]

  \addplot+[only marks,
            mark=*,
            mark options={draw=red, fill=red!20}]
            coordinates {(15,24) (17,23)};
  \addlegendentry{Potential integer solutions}

  \addplot+[only marks,
            mark=*,
            mark options={draw=black, fill=black!20}]
            coordinates {(16,24)};
  \addlegendentry{Rounded smooth solution $\lfloor \hat{\bx}^* \rceil$}

  \addplot+[only marks,
            mark=*,
            mark options={draw=red, fill=red}]
            coordinates {(15.75,23.75)};
  \addlegendentry{Smooth solution $\hat{\bx}^*$}

  \addplot[domain=0:360,samples=200,no marks,thick]
           ({15.75 + 1.25*cos(x) + 0.176*sin(x)},
            {23.75 - 0.75*cos(x) + 0.295*sin(x)});

  \draw[black,dashed,opacity=.85]
        (axis cs:15.75-1.27,23.75-0.81)
        rectangle
        (axis cs:15.75+1.27,23.75+0.81);

  \end{axis}
\end{tikzpicture}
\caption{Exemplary visualization of our proposed method. $\bx_1$ and $\bx_2$ denote integer quantities of two assets. A naive approach would require roughly 5 qubits per dimension to allow for quantities up to $2^5=32$. Our approach constructs an ellipsoid around a smooth solution that can be obtained efficiently via a standard QP solver. Only integer points within the ellipsoid's dashed bounding-box are considered during optimization. Thus, the number of qubits required for each dimension is reduced to 2 and 3, respectively.\label{fig:intro}}
\end{SCfigure}
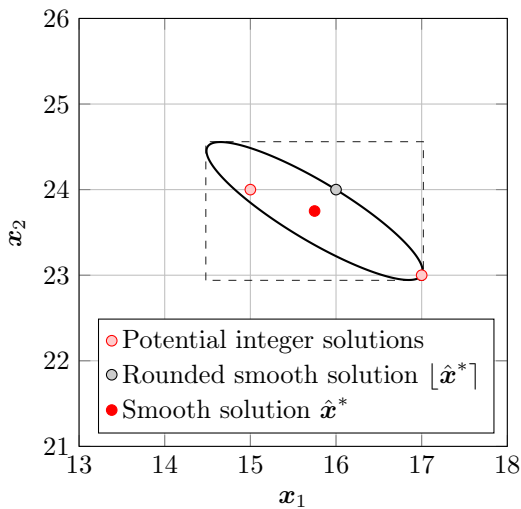

At the core of our work lies the observation that some hard combinatorial optimization problems correspond to a surprisingly simple problem when allowing for continuous solutions. For example, \emph{Quadratic Unconstrained Binary Optimization} (QUBO) with objective $f(\bx)=\bx^\top \bQ \bx$ with $\bx\in\{0,1\}^n$ is NP-hard \cite{Punnen/2022a}. However, when $\bx$ comes from a simplex, the problem becomes a convex \emph{Quadratic Program} (QP) that can be solved to optimality in polynomial time.
Here, this observation is utilized by constructing a new QUBO that incorporates the location of the QP's optimal solution $\bar{\bx}^*\in\mR^n$. This approach allows us to restrict the discrete search space to solutions in the direct neighborhood of $\bar{\bx}^*\in\mR^n$, which results in a lower-dimensional Hilbert-space and thus in a reduced number of qubits when compared to the standard QUBO formulation. The situation is depicted in Fig.~\ref{fig:intro}. 
The search space is endogenously determined to guarantee that it contains the optimum; in general it is not sufficient to round up or down the continuous solution.

Existing quantum approaches to portfolio optimization either model binary inclusion variables per asset
\cite{Egger/etal/2021a,venturelli2019reverse}, 
or assume predefined investment bands of lower and upper bounds per asset \cite{aguilera2024multi,palmer2021quantum,rosenberg2016solving}.
To improve solution quality at limited hardware, papers have used insights to guide the quantum search, for instance, by identifying high-Sharpe-Ratio candidates in a pre-processing step \cite{lang2022strategic} or using the results of a classical heuristic to seed the quantum annealer via reverse annealing \cite{venturelli2019reverse}.
Warm-starting methods using the solution from the continuous relaxation have been explored, including initial state preparation for simulated annealing \cite{rubio2022portfolio} 
and the preparation of quantum states for algorithms on gate-based quantum computers \cite{Egger/etal/2021a}.
To the best of our knowledge, our paper is the first one that integrates the solution of the continuous relaxation into the QUBO construction and thereby endogenously determines relevant investment bands.

We conduct experiments based on real-world S\&P 500 stock market data on software solvers and a D-Wave Advantage quantum annealer. Our results show that our novel formulation can lead to a drastic reduction of qubits---effectively allowing for larger problem instances on actual quantum hardware. 

\section{Background}
\paragraph{Quantum Computing.}

Let us quickly introduce the basic notion of what can be considered as quantum computation \cite{Nielsen/Chuang/2016a}. 
Today, practical quantum computing (QC) consists of two dominant paradigms: Adiabatic QC and gate-based QC. In both scenarios, a quantum state $\ket{\psi}$ for a system with $n$ qubits is a $2^n$ dimensional complex and $\ell_2$-normalized vector. The domain of $\ket{\psi}$ is known as Hilbert space. In the gate-based (universal) framework, a quantum computation is defined as a matrix multiplication $\ket{\boldsymbol{\psi}_{\text{out}}}=C\ket{\boldsymbol{\psi}_{\text{in}}}$, where the $C$ is a $(2^n \times 2^n)$-dimensional unitary matrix (the circuit), given via a 
series of inner and outer products of low-dimensional unitary matrices (the gates). The circuit is derived manually or via heuristic search algorithms \cite{Franken2022}. In adiabatic quantum computation (AQC)---the
framework that we consider in the paper at hand---the result of computation is defined to be the eigenvector $\ket{\phi_{\min}}$ that corresponds to the smallest eigenvalue of some $(2^n \times 2^n)$-dimensional Hermitian matrix $\bH$ (also known as Hamiltonian). In 
practical AQC, $\bH$ is further restricted to be a real diagonal matrix whose entries can be 
written as $\bH_{i,i}=\bH(\bQ)_{i,i}={(\boldsymbol{x}^{(i)})^{\top} \bQ \boldsymbol{x}^{(i)}}$ where $\boldsymbol{x}^{(i)}=\operatorname{binary}(i,n)\in\{0,1\}^n$ is some (arbitrary but 
fixed) $n$-bit binary expansion of the unsigned integer $i$. Here, $\bQ \in \mathbb{R}^{n\times n}$ is the so-called {QUBO} matrix. By construction, computing $\ket{\phi_{\min}}$ is equivalent to solving 
\begin{equation}
	\min_{\boldsymbol{x}\in\{0,1\}^n}\ \boldsymbol{x}^{\top} \bQ \boldsymbol{x}\;.\label{eq:qubo}
\end{equation}Adiabatic quantum algorithms rely on this connection by encoding combinatorial problems into {QUBO} matrices. 

In both paradigms, the output vector is $2^n$-dimensional and can thus not be read-out efficiently for non-small $n$. Instead, the output of a practical quantum computation is a random integer $i$ between 1 and $2^n$, which is drawn from the probability mass function $\mP(i)=|\braket{i | \boldsymbol{\psi}_{\text{out}}}|^2$. This sampling step is also known as collapsing the quantum state $\ket{\boldsymbol{\psi}_{\text{out}}}$ to a classical binary state $\operatorname{binary}(i,n)$.
AQC has been applied to numerous combinatorial optimization problems \cite{lucas2014}, ranging over satisfiability \cite{kochenberger2005}, routing problems \cite{neukart2017} to machine learning \cite{Mucke/etal/2019a}.

The number of physically available qubits of an actual \emph{Quantum Processing Unit} (QPU) is limited. As a result, a theoretically sound QUBO formulation might not be able to run on a real-world QPU as $n$ exceeds the number of available qubits. Decomposition techniques can be applied in these cases, cutting the original QUBO into smaller pieces. It is, however, not guaranteed that re-combining partial solutions will lead to a good (or even feasible) global solution. Hence, another approach is to come up with an alternative QUBO formulation that requires a smaller number of qubits. One example is the integration of inequality constraints via the unbalanced penalty method \cite{Lee/etal/2024a}, which requires less qubits than naive integration of the constraints. In the paper at hand, the QUBO dimension is reduced by reducing the search space to discrete solutions in the vicinity of the optimum of a relaxed convex QP.

\paragraph{Portfolio Theory.}

In the sense of the classical mean-variance approach introduced by \cite{ma52}, 
investors select 
non-negative portfolio weights $\bw$ on a continuous scale, with each entry $\bw_i$ defining the fraction of wealth invested in asset $i$.
Let $\bmu$ denote the expected returns and $\bSig$ the covariance matrix of asset returns.
The selection can be formalized by an optimization problem identifying 
the maximal expected portfolio return subject to a specified level of portfolio variance:
\begin{equation}
\max_{\bw}\ \bmu^\top \bw \text{\quad subject to\quad} \bw^\top \bSig \bw \leq c, \sum_{i=1}^{n} \bw_i = 1, \bw_i \geq 0 \text{ for all } i\;.\label{eq:mvp}
\end{equation}
Problem~\eqref{eq:mvp} is the classical starting point to identify an optimal portfolio with integer amounts of asset purchases.
Let $\bp \in \mR_+^n$ collect the price per unit of each asset, and let $B>0$ denote the budget to be invested.
Similar to \cite{Castro/etal/2011a}, to identify the optimal units of purchased assets $\bx\in \mZ_{+}^n$, one considers the problem
\begin{equation}
\max_{\bx}\ {\tilde \bmu}^\top \bx \text{\quad subject to\quad} \bx^\top \tilde \bSig \bx \leq c \cdot B^2, \sum_{i=1}^{n} \bp_i \cdot \bx_i = B, \bx_i \in \mZ_{+} \text{ for all } i\;.\label{eq:qip0}
\end{equation}
Notice how \eqref{eq:mvp} gives rise to \eqref{eq:qip0} by adding the integrality constraint $\bx_i \in \mZ_{+}$. Their solutions can be translated back and forth via the following equations\footnote{The operator $\circ$ denotes element-wise multiplication.}:
\begin{equation}
\bw = \frac{\bx \circ \bp}{B}, \quad \tilde \bSig = \text{diag}(\bp) \cdot \bSig \cdot \text{diag}(\bp), \quad \tilde \bmu = \bmu \circ \bp \label{eq:convert},
\end{equation}
In order to construct a convex problem over the simplex, the expected portfolio return and the portfolio variance are combined into a quadratic objective function via the risk aversion parameter $\gamma > 0$. 
Consequently, we arrive at
\begin{equation}
\max_{\bw}\ \bmu^\top \bw - \frac{\gamma}{2} \cdot \bw^\top \bSig \bw \text{\quad subject to\quad} \sum_{i=1}^{n} \bw_i = 1, \bw_i \geq 0 \text{ for all } i\;.\label{eq:QP}
\end{equation}

In \eqref{eq:qip0}, it would be restrictive to require that the budget is invested exactly, $\sum_{i=1}^{n} \bp_i \cdot \bx_i = B$.
Instead, we assume that the difference between the invested amount and the budget $B$ is invested at the risk-free interest rate $r_f$. 
In this case, the $\bw_i$ represent the weights of each risky asset, while $\left(1-\sum_{i=1}^{n} \bw_i \right)$ represents the weight to be invested at the risk-free rate. 
Consequently, the constraint $\sum_{i=1}^{n} \bw_i = 1$ can be removed and the
expected return becomes 
\begin{equation}
\bmu^\top \bw + \left( 1- \bone^\top \bw \right) \cdot r_f = 
\left(\bmu - r_f \cdot \bone \right)^\top \bw + r_f
\end{equation}
with $\bone$ denoting an $n$-dimensional vector of ones. 
Moreover, we omit the no-shortselling constraint $\bw_i\geq 0$, as in practice transaction costs may imply that the investor slowly trades from an existing portfolio to the preferred one. 
Therefore, the optimal solution typically remains long-only, so the constraint is empirically non-binding (as seen in Sec.~\ref{sec:experiments}).
The new problem can be formulated as
\begin{equation}
\max_{\bw}\ {\underbrace{\left(\bmu - r_f \cdot \bone \right)}_{\bmu_f}}^\top \bw  - \frac{\gamma}{2} \cdot \bw^\top \bSig \bw\;. 
\label{eq:QP2}
\end{equation}
Setting $\tilde \gamma = \gamma/B$ and invoking Eq.~\ref{eq:convert} for $\bmu_f$, we arrive at
\begin{equation}
\max_{\bx}\ \tilde \bmu_f^\top \bx - \frac{\tilde \gamma}{2} \cdot \bx^\top \tilde \bSig \bx \text{\quad subject to\quad} \bx_i \in \mZ \text{ for all } i\;.\label{eq:QP3}
\end{equation}

The latter represents the NP-hard portfolio \emph{Quadratic Integer Program} (QIP). 
However, since $\bSig$ is a covariance matrix which is typically positive definite, dropping the integrality constraint of \eqref{eq:QP3} leads to a smooth convex quadratic program for which solutions can be readily obtained via standard solvers.
For empirical situations with sample covariance matrices being ill-conditioned or singular, regularization methods are widely used 
to ensure positive definiteness; see, e.g., \cite{ledoit2008robust}.
Furthermore, the structural form of the mean-variance framework in \eqref{eq:QP3} can be derived, at least approximately, for homogeneous risk measures such as Value-at-Risk or Expected Shortfall, as well as for deterministic regulatory capital standards (see \cite{pasc22}).
Even when \eqref{eq:QP3} is extended to include additional linear or convex quadratic terms, such as transaction costs, a smooth convex quadratic program can be established (see Sec.~\ref{sec:experiments}).

In order to construct an equivalent QUBO, $\bx$ must be binarized, e.g., by reserving a fixed amount of $k$ bits for each asset and interpreting the bits as unsigned binary expansion of some integer. Clearly, the choice of $k$ has an immediate impact on both, the accuracy to which we can represent the asset weights and the dimension of the resulting QUBO. We will re-visit this topic in our experiments.
\section{Methodology} \label{sec:method}
%
By adopting ideas from \cite{Buchheim/etal/2012a}, we derive a novel method to reduce the number of qubits required for the portfolio optimization QUBO. 
The proposed method relies on an informed re-formulation of the portfolio QIP.

\begin{theorem}[Informed Ellipsoid Formulation]\label{thm:informed}
Let ${\bx}^*$ be an optimal solution to \eqref{eq:QP3} with $\tilde \bSig$ positive definite, 
$\hat{\bx}^*$ the smooth maximizer of $f(\bx)=\tilde \bmu_f^\top \bx - \frac{\tilde \gamma}{2} \cdot \bx^\top \tilde \bSig \bx$, and $\hat{f}^*:=f(\hat{\bx}^*)$ the corresponding optimal function value.
Assume that at least one entry of $\hat{\bx}^*$ is not integer-valued.
Let further
\begin{equation}
\bE := \left\{ \bx \in \mR^n : \left(\bx-\hat{\bx}^*\right)^\top  \left( \frac{\tilde \gamma}{2C} \cdot \tilde \bSig \right)  \left(\bx-\hat{\bx}^*\right) \leq 1 \right\}
\label{def:E}
\end{equation}
with $C=\hat{f}^* - f(\lfloor \hat{\bx}^* \rceil)>0$. Then, $\bx^* \in \bE$. Thus, optimizing over the ellipsoid $\bE$ is sufficient to compute $\bx^*$. Here, $\lfloor \cdot \rceil$ denotes component-wise rounding to the nearest integer. 
\end{theorem}
\textbf{Proof}. First, notice that 
\begin{equation*}
\begin{split}
\left(\bx-\hat{\bx}^*\right)^\top  \left( \frac{\tilde \gamma}{2}  \tilde \bSig \right)  \left(\bx-\hat{\bx}^*\right) \leq C 
&\Leftrightarrow\ 
 \left(\bx-\hat{\bx}^*\right)^\top  \left( \frac{\tilde \gamma}{2}  \tilde \bSig \right)  
\left(\bx-\hat{\bx}^*\right) - \hat{f}^* \leq -f(\lfloor \hat{\bx}^* \rceil)\\
&\Leftrightarrow\ 
 (\hat{\bx}^*)^\top  \left( \tilde \gamma  \tilde \bSig \right)  \bx - \frac{\tilde \gamma}{2}  \bx^\top  \tilde \bSig  \bx \geq f(\lfloor \hat{\bx}^* \rceil)\;.
\end{split}
\end{equation*}
Moreover, since $\hat{\bx}^*$ satisfies the first order condition of $f$, it follows that
\begin{equation*}
 \tilde\bmu_f - \tilde \gamma  \tilde\bSig \hat{\bx}^* = 0 \Leftrightarrow \hat{\bx}^* = \frac{1}{\tilde \gamma}  \tilde\bSig^{-1}  \tilde\bmu_f\;.
\end{equation*}
Plugging this into the last inequality yields
\[
\tilde\bmu_f  \bx - \frac{\tilde \gamma}{2}   \bx^\top  \tilde \bSig  \bx = f(\bx)  \geq f(\lfloor \hat{\bx}^* \rceil)\;.
\]
We see that all integer points contained in $\bE$ (if any) achieve a better objective function value than the rounded smooth solution $\lfloor \hat{\bx}^* \rceil$. Since all points outside of $\bE$ are strictly worse than $\lfloor \hat{\bx}^* \rceil$, the best integer solution must be included in $\bE$ as well.
\ \hfill$\blacksquare$

Based on these insights, we compute the extent of $\bE$ for each asset $i$ in terms of upper and lower bounds. 
The lower and upper boundaries of the ellipsoid in all dimensions can then be computed via
\[
L = \hat{\bx}^* - \sqrt{\operatorname{diag}(\bM^{-1})}
\text{\quad and\quad}
U = \hat{\bx}^* + \sqrt{\operatorname{diag}(\bM^{-1})}
\]
with $\bM= \frac{\tilde \gamma}{2C} \cdot  \tilde \bSig $.
Finally, instead of optimizing over $\bx\in\mZ^n$, we exploit the limits of the ellipsoid and optimize over 
\begin{equation}
\bx_i\in[L_i,U_i]\cap\mZ \text{ for all } i
\label{def:cube}
\end{equation}
This corresponds to optimization over a bounding hypercube that tightly encloses the ellipsoid\footnote{While this introduces sub-optimal solutions into the search space, it does not harm the soundness of our result. Nevertheless, the effectiveness of our approach can be enhanced through several methods; for example, by successively refining the search space based on the best integer solution found, and/or by including penalties in the objective function to more tightly bound the search space.} (see Fig.~\ref{fig:intro}). 
Since $[L_i,U_i]$ can be different for each $i$, the effective search space is adapted to the individual extent of each dimension.
Thus,  this procedure has the potential to reduce the search space, and hence the number of required qubits, by a large margin. We call this procedure \emph{hot-start}.




\section{Experiments}\label{sec:experiments}
Two experiments are conducted to demonstrate the viability of our approach. 
First, the number of qubits required by a baseline $k$-bits per asset construction is compared to the number of qubits required for the proposed approach for asset universes of increasing size. Second, we compare the solution quality obtained on a D-Wave Advantage Quantum Annealer and a randomized search heuristic, both utilizing the baseline and our novel construction. 

We calibrate $\tilde\bmu$ and $\tilde\bSig$ using monthly S\&P 500 equity returns from August 2003 to July 2023, sourced from Refinitiv.  
The vector $\bmu$ and the matrix $\bSig$ are estimated as the sample mean and sample covariance matrix of the stock returns over this period.
The vector $p$ contains stock prices as of July 2023; the risk-free interest rate is set to $r_f = 5.32\%$ based on the Federal Funds Effective Rate, as reported by the Federal Reserve Bank of St. Louis (FRED).

Our investment universe consists of the $n$ stocks with the highest market capitalization as of July 2023.  
The investor has a budget of $B = 250{,}000\,\text{USD}$ and initially holds a portfolio with (approximately) equal weighting across all $n$ assets in the investment universe.
The initial portfolio is integer-valued, i.e., rounded as
\begin{equation}
    \bx_0 = \left\lfloor \frac{B}{n} \cdot \bp^{-1} \right\rfloor
\end{equation}
where $\bp^{-1}$ denotes the vector of element-wise reciprocals of the price vector $\bp$
and $\lfloor \cdot \rfloor$ denotes component-wise flooring.
We consider problem \eqref{eq:QP3}, supplemented by transaction costs.
As in \cite{garleanu2013dynamic}, we assume that transaction costs are determined by the covariance matrix $\bSig$ and thus achieve
\begin{equation}
\max_{\bx}\ \tilde \bmu_f^\top \bx - \frac{\tilde \gamma}{2} \cdot \bx^\top \tilde \bSig \bx
 - \tilde \kappa \cdot \left(\bx - \bx_0 \right)^\top \tilde \bSig \left(\bx - \bx_0 \right)
\text{\quad subject to\quad} \bx_i \in \mZ \text{ for all } i\;.
\label{prob:example}
\end{equation}
We set $\gamma = 3$ and $\tilde{\kappa} = 50 \cdot \gamma / B$.\footnote{$\gamma=3$ lies within the range of economically realistic and empirically estimated risk-aversion coefficients; for a meta-analysis see \cite{elminejad2022relative}.
Compared to empirical estimates, our calibration of transaction costs is moderate; for example, \cite[p.~2328]{garleanu2013dynamic} estimate $\kappa$ for several commodities and find that, in the median, $\kappa$ is a factor of 500 larger than $\gamma$.}

For the $n=4$ stocks with highest market capitalization, Table~\ref{tab:4assets} shows how Hot-Start 
constructs a QUBO with six qubits. Starting from the initial portfolio $\bx_0$, 
the smooth solution 
$\hat{\bx}^*$
of \eqref{prob:example} suggests reducing positions 
in Microsoft, NVIDIA, and Amazon, while increasing the position in Apple.
Enclosing hypercube \eqref{def:cube} around ellipsoid $\bE$ from \eqref{def:E} 
implies that the optimal integer-valued solution of \eqref{prob:example} must be 
either 194 or 195 for Microsoft, either 372 or 373 for Apple, and so on. 
Thus, \eqref{prob:example} can be solved as a QUBO requiring 
$1+1+3+1=6$ qubits, where the number of qubits is determined by 
the binary encoding rule 
$\#\text{qubits} = \left\lceil \log_{2}\!\bigl(\#\text{integers}\bigr) \right\rceil$ 
for each stock.

Increasing $n$ successively to 100, 
the maximum number of qubits required by Hot-Start for any single asset increases from 3 (for $n=4$) to 10 (for $n=100$). 
To provide a conservative baseline for comparison, we calibrate a fixed $k$-bits per asset construction with $k=10$.
This baseline is as efficient as possible while still ensuring that the optimal solution is not overlooked.

\begin{SCfigure}[][t]
\centering
\begin{tikzpicture}
  \begin{axis}[
    width=9cm,
    height=6cm,
    xlabel={Number of Stocks},
    ylabel={Number of Qubits},
    grid=both,
    xtick={0,20,40,60,80,100},
    ytick={0,200,400,600,800,1000},
    ymin=0,
    ymax=1005,
    enlarge x limits=0.05,
    enlarge y limits=0.05,
    line width=1pt,
    legend pos=north west,
  ]
    \addplot[
      color=blue,
      mark=*,
      mark options={fill=blue},
    ] coordinates {
      (  4,   6)
      ( 10,  16)
      ( 20,  54)
      ( 40, 154)
      ( 60, 362)
      ( 80, 487)
      (100, 599)
    };
    \addlegendentry{Hot-Start (ours)}

    \addplot[
      color=red,
      mark=square*,
      mark options={fill=red},
    ] coordinates {
      (  4,   40)
      ( 10,  100)
      ( 20,  200)
      ( 40,  400)
      ( 60,  600)
      ( 80,  800)
      (100, 1000)
    };
    \addlegendentry{$k=10$ qubits per Asset}
  \end{axis}
\end{tikzpicture}
\caption{Number of qubits required to encode the portfolio QUBO, as a function of the number of stocks. Comparison between our approach (red boxes) and a baseline (blue circles) where each stock quantity is encoded by $10$ qubits.\label{fig:qubit_scaling}}
\end{SCfigure}
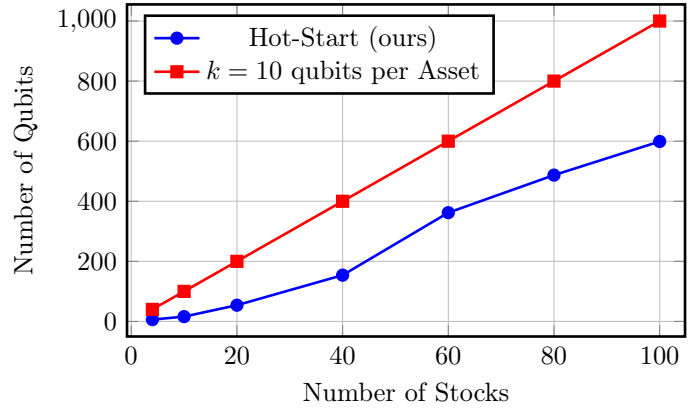

Fig.~\ref{fig:qubit_scaling} shows that in the first experiment, Hot-Start requires a substantially smaller number of qubits than the baseline. 
For the second experiment, we consider the D-Wave Advantage2 Prototype 2.6 QPU with 128 readouts (shots) per QUBO. Before sending the QUBO to the QPU, it must be mapped onto the Zephyr hardware topology. In case of the baseline, this step failed for all problems with more than 40 stocks. In case of Hot-Start, all problems with up to 80 stocks could be mapped successfully to the QPU. The energies of the found solutions coincide with those found by the D-Wave simulated annealing software solver. We refer to \cite{Mucke/etal/2025a} for a comprehensive discussion on how to reduce integrated control errors and thus hardware noise on D-Wave annealers. 

\begin{table}[ht]
\centering
\setlength{\tabcolsep}{7pt} 
\caption{Derivation of the number of qubits for problem \eqref{prob:example} with $n=4$ stocks.}
\begin{tabular}{lrrrrrrr}
\toprule
Stock & $\bx_0$ & $\hat{\bx}^*$ & \multicolumn{2}{c}{Integers in box} & \multicolumn{2}{c}{Decision variables} \\
\cmidrule(lr){4-5} \cmidrule(lr){6-7}
 & & & Smallest & Largest & \# integers & \# qubits \\
\midrule
Microsoft & 197  & 194.6  & 194  & 195  & 2 & 1 \\
Apple     & 365  & 372.6  & 372  & 373  & 2 & 1 \\
NVIDIA    & 1436 & 1434.2 & 1432 & 1436 & 5 & 3 \\
Amazon    & 491  & 489.7  & 489  & 490  & 2 & 1 \\
\midrule
Total     &      &        &      &      &  &  6 \\
\bottomrule
\end{tabular}
\label{tab:4assets}
\end{table}

\section{Concluding Remarks}
We presented Hot-Start, the first warm-start technique that lowers the qubit demand for quantum portfolio optimization. Experiments show favorable scaling: the conventional encoding lacks prior knowledge of the qubit count needed to represent a solution and therefore over-allocates, whereas Hot-Start infers that number automatically and uses qubits more efficiently. With Hot-Start we successfully embedded QUBOs for portfolios of up to 80 stocks on current QPUs, while the baseline approach already failed beyond 40 stocks. Because the approach relies only on the existence of a convex QP representation, we expect Hot-Start to transfer to a broad class of optimization problems well beyond finance, providing an effective pathway for harnessing near-term and future QPUs.

\subsection*{Acknowledgment}
This research has been funded by the Federal Ministry of Education and Research of Germany and the state of North-Rhine Westphalia as part of the LAMARR Institute for Machine Learning and Artificial Intelligence.

\bibliographystyle{splncs04}
\bibliography{main,references,refs}

\end{document}